\documentclass[twocolumn,superscriptaddress,showpacs,aps,floatfix,nofootinbib]{revtex4-1}

\usepackage[english]{babel}

\usepackage{soul}
\usepackage{orcidlink}
\usepackage{comment}
\usepackage{dcolumn} 

\usepackage{amsmath}
\usepackage{amsfonts}
\usepackage{bm}
\usepackage{siunitx}
\usepackage{nicefrac}
\usepackage{physics}

\usepackage{array}
\usepackage{multirow}

\usepackage{graphicx}
\usepackage{xcolor}
\usepackage{xspace}
\usepackage{float}

\usepackage{hyperref}

\newcommand{\vinca}{\affiliation{Vin\v{c}a Institute of Nuclear Sciences - 
National Institute of the Republic of Serbia, University of Belgrade, P. O. Box 522, RS-11001 Belgrade, Serbia}}
\newcommand{\cnrspin}{\affiliation{Consiglio Nazionale delle Ricerche CNR-SPIN, c/o Universit\'{a} degli Studi "G. D’Annunzio", 66100 Chieti, Italy}}

\newcommand{\pmf}{\affiliation{Faculty of Natural Sciences and Mathematics, University of Montenegro, D\v{z}ord\v{z}a Va\v{s}ingtona bb, 81000 Podgorica, Montenegro}}
\newcommand{\bicocca}{\affiliation{Department of Materials Science, University of Milan-Bicocca, Via Roberto Cozzi 55, 20125 Milan, Italy}}
\newcommand{\ua}{\affiliation{Department of Physics and NANOlab Center of Excellence, University of Antwerp, Groenenborgerlaan 171, B-2020 Antwerp, Belgium
}}%

\begin{document}

\title{Hole doping as an efficient route to increase the Curie temperature in monolayer CrI$_3$}

\author{Marko Orozovi\'c \orcidlink{0000-0002-3018-5830}} \vinca
\author{Bo\v{z}idar N. \v{S}o\v{s}ki\'c \orcidlink{0000-0003-3482-7359}} \pmf \ua
\author{Silvia Picozzi} \bicocca \cnrspin
\author{\v{Z}eljko \v{S}ljivan\v{c}anin \orcidlink{0000-0001-8575-2575}} \vinca
\author{Srdjan Stavri\'c \orcidlink{0000-0003-2097-0955}}\email{stavric@vin.bg.ac.rs} \vinca

\date{\today} 

\begin{abstract}
\noindent
Two-dimensional van der Waals (vdW) magnets offer unprecedented opportunities to control magnetism at the atomic scale. Through charge carrier doping---realized by electrostatic gating, intercalation/adsorption, or interfacial charge transfer---one can efficiently tune exchange interactions and spin-orbit-induced effects in these systems. In this work, through a multi-scale theoretical framework combining density functional theory, spin Hamiltonian modeling, and Wannier-function analysis, we choose monolayer CrI$_3$ to unravel how carrier doping affects the isotropic as well as anisotropic exchange interactions in this prototypical vdW ferromagnet. The remarkable efficiency of hole doping in enhancing ferromagnetic exchange and magnetic anisotropy found in our study was explained through orbital-resolved analysis. Crucially, we demonstrated that unlike the undoped system—where isotropic exchange interactions govern magnetic long-range order—the hole-doped CrI$_3$ exhibits anisotropic terms comparable in magnitude to isotropic ones. Finally, we show that a high concentration of holes in a CrI$_3$ monolayer can increase its Curie temperature above 200 K. This work advances our understanding of doping-controlled magnetism in semiconducting 2D materials, demonstrating how anisotropy engineering can stabilize high-temperature magnetic order.
\end{abstract}


\maketitle

\section{Introduction}

The discovery of ferromagnetic order in the ultrathin limit of layered van der Waals (vdW) materials CrI$_3$~\cite{Huang2017Jun} and CrGeTe$_3$~\cite{Gong2017Jun} ignited a decade of intense research on two-dimensional (2D) magnets. Building on these initial discoveries, recent studies estimate that more than 800 vdW materials may exhibit long-range magnetic order down to the monolayer limit~\cite{Gjerding2021Jul}. Among these, more than two dozen have been experimentally synthesized so far~\cite{Park2025May}. 
Besides ferromagnetism (FM) and antiferromagnetism (AFM), vdW materials can host altermagnetism~\cite{Milivojevic2024May,Sodequist2024May} as well as various noncollinear magnetic textures like skyrmions~\cite{Amoroso2020Nov,Zhang2022Mar}, merons~\cite{Lu2020Oct,Augustin2021Jan} or both~\cite{Casas2023Apr}.
Furthermore, multiferroic vdW materials that have coupled magnetic and ferroelectric properties enable electrical control of their magnetic properties, such as that recently achieved with $p$-wave magnetic ordering in type-II vdW multiferroic NiI$_2$~\cite{Song2022Feb,Song2025Jun}.
Therefore, having at our disposal ultrathin vdW materials that can host various kinds of magnetic phases and textures, as well as those with cross-coupled magnetic and ferroelectric properties, the time seems ripe for all-2D spintronics devices~\cite{Zhong2025Feb,Gong2025May}.

The realization of practical 2D spintronic devices requires 2D crystals that exhibit stable magnetic order at temperatures exceeding room temperature.
In fact, several 2D magnets already satisfy this requirement. For example, room-temperature ferromagnetism is confirmed by the observation of magnetic hysteresis at $330 \, {\rm K}$ in ${\rm VSe_2}$~\cite{Bonilla2018Apr} and at $300 \, {\rm K}$ in ${\rm MnSe_2}$~\cite{O'Hara2018May} monolayers.
In a Cr$_3$Te$_4$ monolayer, x-ray magnetic circular dichroism measurements revealed a critical temperature $(T_{\rm c})$ of $344 \, {\rm K}$ with an out-of-plane magnetic easy axis~\cite{Chua2021Oct}. The $T_{\rm c}$ of a Fe$_3$GaTe$_2$ 2D crystal was estimated at $\sim 350 –380 \, {\rm K}$, which was at a time a record-high $T_{\rm c}$ for intrinsic 2D vdW ferromagnetic crystals. However, the room-temperature $T_{\rm c}$ values are exclusively reported for itinerant 2D ferromagnets, whereas semiconducting 2D ferromagnets exhibit significantly lower $T_{\rm c}$. For instance,  the widely studied CrI$_3$ and CrGeTe$_3$ have $T_c$ values of 45 and 42 K, respectively.

To enhance the $T_c$ in 2D magnetic semiconductors and especially in CrI$_3$, various strategies have been proposed in recent years. In the work of Zhang \textit{et al.}~\cite{Zhang2021Mar} it is argued that defect engineering---creating Cr and I atomic vacancies in CrI$_3$---should result with a 2D system possessing higher $T_{\rm c}$, although no concrete $T_{\rm c}$ value has been reported. Chen \textit{et al.} have predicted that CrI$_3$/MoTe$_2$ heterostructure should exhibit $T_{\rm c} \sim 60 \, {\rm K}$, with further increase up to $85 \, {\rm K}$ that can be reached by decreasing interlayer distance~\cite{Chen2019Jul}. 
Surface functionalization has also been proposed as an efficient means to increase $T_{\rm c}$ in 2D magnets. For instance, adsorption of halogen atoms (F, Cl, Br) on CrI$_3$ is predicted to increase $T_{\rm c}$ up to $\sim 220 \, {\rm K}$~\cite{Li2021Mar}; the Li adsorption can lead to $T_{\rm c} \sim 150 \, {\rm K}$ at $12.5\%$ Li coverage~\cite{Xu2020Oct}; the substitutional doping of I with Se impurities can increase the $T_{\rm c}$ up to $\sim 250 \, {\rm K}$; the adsorption of transition metals Sc and V can lead to a nearly threefold increase in $T_{\rm c}$~\cite{Yang2021May}. 

Among the suggested strategies to increase the $T_{\rm c}$ in CrI$_3$ \textit{charge carrier doping} has emerged as a particularly promising one. 
Wang \textit{et al.}~\cite{Wang2016Jun} predicted that hole doping raises $T_{\rm c}$ by introducing itinerant carriers that strengthen the FM exchange between localized Cr magnetic moments. While their calculations also suggested a modest $T_{\rm c}$ enhancement under electron doping, the effect was predicted to be significantly weaker than in the hole-doping counterpart.
Experimental validation of these predictions was provided by Jiang \textit{et al.}~\cite{Jiang2018Jul}, who demonstrated that electrostatic gating of a single CrI$_3$ layer enhances $T_{\rm c}$ by 20\% at a concentration of $\sim 0.11$ holes per unit cell. In contrast, electron doping was found to suppress $T_{\rm c}$---a result that contradicts initial theoretical expectations. First-principles calculations by Singh \textit{et al.}~\cite{Singh2021Jun} resolved this discrepancy by showing that electron doping induces tensile strain in CrI$_3$, counteracting the electronic enhancement of ferromagnetic exchange and leading to an overall suppression of exchange interactions at higher doping concentrations. Moreover, the calculations by Singh \textit{et al.} reproduce the significant enhancement of $T_c$ in hole-doped CrI$_3$ and pinpoint key contributing factors. However, a full microscopic explanation is still lacking---particularly regarding the stark electron-hole doping asymmetry.

In this theoretical study, we present a comprehensive investigation of electron- and hole-doping effects on the electronic and magnetic properties of monolayer CrI$_3$, with particular emphasis on their impact on $T_{\rm c}$. Through orbital-resolved analysis of exchange interactions, we identify the specific interaction channels affected by carrier doping. Given the critical role of magnetic anisotropy in stabilizing long-range order in 2D magnets~\cite{Mermin1966Nov}, we systematically evaluate the doping-induced changes in single-ion anisotropy (SIA) and in both the isotropic and anisotropic exchange-terms beyond nearest neighbors, which increase substantially when itinerant carriers are introduced in a semiconducting CrI$_3$ monolayer. Additionally, we compute the doping-induced changes in magnetic anisotropy energy (MAE), which encompasses all anisotropic contributions, and demonstrate why, in the case of hole doping, the SIA counteracts the overall increase in MAE. Our calculations explain the superior efficiency of hole doping over electron doping for enhancing the ferromagnetic exchange and magnetic anisotropy in the CrI$_3$ monolayer and consequently increasing $T_{\rm c}$, which is of both fundamental and technological significance.  

\section{Computational Methodology}
\subsection{Model spin Hamiltonian}
\label{sseq:hamiltonian}

To study the magnetic properties of ${\rm CrI_3}$ monolayer, we use a generalized model of the bilinear spin Hamiltonian,
\begin{equation}
\label{eq:H}
    \mathcal{H} = -\sum^{_{1,2,3}}_{i<j} {\bf S}_i \mathcal{J}_{ij} {\bf S}_j - \sum_i {\bf S}_i \mathcal{A}_{ii} {\bf S}_i.
\end{equation}
The spins $ {\bf S}_i = (S_{i,x},S_{i,y},S_{i,z})$ are classical pseudovectors with magnitude $S = 3/2$ distributed over honeycomb lattice sites $i$. We adopt the convention where positive (negative) values of the exchange parameter $J$ correspond to FM (AFM) coupling. The "$1,2,3$" symbol above the first sum in equation~(\ref{eq:H}) denotes that the summation is over all nearest neighbors (NN) including the third. These neighbors are depicted in figure~\ref{fig:isotropic}(a). We omit exchange interactions with fourth and more distant neighbors due to their nearly negligible contributions, as shown in section IV of the supplementary information. 

The $3\times3$ matrix $\mathcal{J}$, representing generalized exchange interaction between two selected spins, decomposes into three components using the standard symmetrization procedure~\cite{Li2021Feb}, 
\begin{equation}
\label{eq:J}
\begin{aligned}
\mathcal{J} &= \underset{\text{isotropic exchange}}{\frac{1}{3}{\rm Tr}({\mathcal{J}}) \mathbb{I}_3}
+ \underset{\text{DMI}}{\frac{1}{2}(\mathcal{J}-\mathcal{J}^T)} \\
&\quad + \underset{\text{anisotropic symmetric exchange}}{
\left[\frac{1}{2}(\mathcal{J}+\mathcal{J}^T)-\frac{1}{3}{\rm Tr}(\mathcal{J})\mathbb{I}_3\right]} \\
 &= J\mathbb{I}_3 + \mathcal{D} + \mathcal{K}.
\end{aligned}
\end{equation}
The first component $J\mathbb{I}_3$ represents the Heisenberg isotropic exchange term and is the only component of the $\mathcal{J}$ matrix that is independent of the spin-orbit coupling (SOC). 

Bilinear form ${\bf S}_i\mathcal{D}{\bf S}_j$ corresponds to the Dzyaloshinskii-Moriya interaction (DMI), which is usually expressed in the form ${\bf D} \cdot ({\bf S}_i\times{\bf S}_j)$, where the components of the vector ${\bf D} = (D_x, D_y, D_z)$ are related to the antisymmetric part of the exchange tensor $\mathcal{J}$ via $D_x = \frac{1}{2}(J_{yz} - J_{zy})$, $D_y = \frac{1}{2}(J_{zx} - J_{xz})$, and $D_z = \frac{1}{2}(J_{xy} - J_{yx})$. In a CrI$_3$ monolayer, according to Moriya's symmetry rules~\cite{Moriya1960Oct}, the DMI is forbidden between first and third NN due to the presence of inversion symmetry at the Cr-Cr bond center, but is allowed between second NN. However, most computational studies on CrI$_3$ completely neglect this DMI as its magnitude is significantly smaller than that of the isotropic exchange~\cite{Kvashnin2020Sep,Vishkayi2020Sep}. Yet, as we show later on, such an assumption does not hold any longer in the hole-doped CrI$_3$. 

The last term in equation~(\ref{eq:J}) represents the anisotropic symmetric exchange, arising from crystal symmetry and SOC, which favors spin alignment along specific crystallographic directions through spin-spin interaction. The preferred spin directions are determined by the eigenvectors of the symmetric traceless matrix $\mathcal{K}$\footnote{We use the letter $\mathcal{K}$ to label this matrix as it is related to the Kitaev exchange, but we emphasize that $\mathcal{K}$ \textit{is not} the Kitaev matrix. For instance, the Kitaev matrix can have a non-zero trace.}. While simplified spin models take into account only the $K_{zz}$ component of this matrix, resulting in the widely used XXZ model~\cite{Lado2017Jun,Singh2021Jun}, we instead retain the full $\mathcal{K}$ matrix in our analysis.

The second term in equation~(\ref{eq:H}) represents the SIA, which defines the preferred spin orientation independently of spin–spin interactions. The corresponding on-site interaction is often expressed as a symmetric SIA matrix $\mathcal{A}$. In ${\rm CrI_3}$ monolayer, symmetry constraints reduce this matrix to a single nonzero component $A_{zz}$~\cite{Sabani2020Jul}, signaling that the system prefers the out-of-plane (in-plane) orientation of each spin for $A_{zz}>0$ ($A_{zz}<0$). After these considerations, we arrive at the final form of the spin Hamiltonian that we use in this work, 
\begin{equation}
\begin{aligned}
    \mathcal{H} = &-\sum_{i<j}^{_{1,2,3}}J_{ij}{\bf S}_i \cdot {\bf S}_j 
    - \sum^{_{1,2,3}}_{i<j} {\bf S}_i \mathcal{K}_{ij} {\bf S}_j \\ &- \sum_{i<j}^{_{2}}{\bf D}_{ij} \cdot ({\bf S}_i\times{\bf S}_j) - \sum_i A_{zz}(S_{i,z})^2.
\end{aligned}
\label{eq:H_final}
\end{equation}

We note that similar bilinear spin models have already been employed in previous studies of ${\rm CrI_3}$, though with varying approximations and omissions of certain terms. For example, Lado and Fernández-Rossier~\cite{Lado2017Jun} use XXZ model limited to first NN interactions. Similarly, Xu \textit{et al}.~\cite{Xu2018Nov} restrict $\mathcal{H}$ to first NN interactions; however, they start from the tensor form of $\mathcal{H}$ (as we do here) and transform it into the local coordinate frames of Cr--Cr pairs, where the Kitaev interaction can be described more transparently. Singh \textit{et al.}~\cite{Singh2021Jun} consider an XXZ model that includes isotropic and some anisotropic interactions\footnote{The spin Hamiltonian they use contains the $\Lambda_k S_i^z S_j^z$ term that describes the two-site anisotropic exchange between the $z$-component of spins.} beyond the first NN yet entirely neglect the DMI. 
As we demonstrate in the remainder of this work, both the DMI and the anisotropic symmetric exchange $\mathcal{K}$ beyond the first NN are significantly enhanced when itinerant carriers are introduced in CrI$_3$, reaching energy scales on par with isotropic exchange interactions. 

\subsection{Computation of exchange parameters}

Density functional theory (DFT) calculations were performed using Vienna ab initio simulation package (VASP)~\cite{Kresse1996Oct}. Exchange and correlation effects were described at the generalized gradient approximation (GGA) level using the Perdew-Burke-Ernzerhof (PBE) functional~\cite{Perdew1996Oct}. An energy cutoff for plane wave basis set expansion of the Kohn-Sham wavefunctions used in calculations was set to $350 \, \text{eV}$. We treated Cr $3p, 3d, 4s $ and I $5s, 5p$ as valence states with projector augmented wave (PAW) pseudopotentials~\cite{Blochl1994Dec}. A criterion for self-consistency was set to $10^{-6} \, \text{eV}$. The lattice constant of the CrI$_3$ monolayer $a = 7.005 \, \text{\AA}$ was obtained from non-relativistic spin-polarized DFT calculations in the FM ground state. The atomic positions were relaxed until the maximal force on each atom was less than $ 0.005 \, \text{eV\AA}^{-1}$. We used $22 \, \text{\AA}$ vacuum space to separate the periodic replicas along the $z$-axis. The Brillouin zone (BZ) was sampled using an $8 \times 8 \times 1$ Monkhorst-Pack mesh~\cite{Monkhorst1976Jun}, corresponding to a $k$-point density of approximately $0.015 \, \text{\AA}^{-2}$.

Within the DFT approach carrier doping is realized by varying the total number of electrons in the system and adding a uniform background potential to maintain charge neutrality. We investigate doping concentrations ranging from 1 hole to 0.5 electrons per unit cell, significantly exceeding the experimentally achieved doping level of 0.11 electrons via electrostatic gating~\cite{Jiang2018Jul}. Throughout this work, all doping concentrations are specified per unit cell (u.c.), which contains two Cr atoms. For each doping concentration we rigorously verify the absence of charge leakage into vacuum, as exposed in the supplementary information and illustrated in figure S1.  We maintain the undoped CrI$_3$ atomic positions for all doping concentrations to isolate purely electronic effects on magnetic exchange interactions. This approach allows us to: (i) directly correlate changes in orbital occupation with modifications to exchange pathways, (ii) eliminate confounding strain-induced contributions, and (iii) test Anderson's superexchange theory~\cite{Anderson1959Jul} for doped systems by preserving the original structure, which is essential for such analysis.   

Exchange and SIA tensors, $\mathcal{J}_{i}$ and $\mathcal{A}$, are calculated with non-collinear DFT calculations with SOC included and using the four-state method~\cite{Xiang2012Dec,Li2021Feb}.These results are presented in tables~S2 and S3 of the supplementary information. To calculate $\mathcal{J}_1$ and $A_{zz}$ we used the $2\times2\times1$ supercell and sampled its BZ with $4\times4\times1$ $k$-points. Calculation of longer-distance exchange tensors $\mathcal{J}_2$ and $\mathcal{J}_3$ required larger $3\times3\times1$ supercell to exclude artificial interaction with periodic replicas. The BZ of larger supercell was sampled with $2\times2\times1$ $k$-points.

In addition to four-state calculations, isotropic exchange parameters $J_i$ are calculated with Green's function method developed by Liechtenstein \textit{et al}.~\cite{Liechtenstein1987May}, treating infinitesimal local spin rotations as a perturbation, as implemented in the \texttt{TB2J} code~\cite{He2021Jul}. Given that this method requires the use of a localized basis set, we turned to the \texttt{WANNIER90} code~\cite{Pizzi2020Jan} to construct maximally localized Wannier functions. For the initial guesses of these functions, we used $d$-like orbitals for Cr and $p$-like orbitals for I atoms. To simulate the doping within \texttt{TB2J}, we employed the rigid-band approximation by shifting the Fermi level of the undoped system and using the shifted Fermi level as an upper bound for the Green's function integration. This allows us to capture doping-induced changes in magnetic interactions without recalculating the Wannier functions for each doping concentration.

Curie temperatures were calculated using the Monte-Carlo (MC) Metropolis algorithm implemented in the VAMPIRE software package~\cite{Evans2014Feb}. Simulations were performed using the generalized Heisenberg spin Hamiltonian defined in equation~(\ref{eq:H_final}). For MC simulations we used the rhombohedral unit cell, as defined in section II of the supplementary information, and created a supercell with a size of $700  \, {\rm \AA} \times 700 \, {\rm \AA}$ that accommodates $ 23200 $ spins. Thermal equilibrium was reached after $ 4 \times 10^4 $ steps, followed by an additional $ 4 \times 10^4 $ steps for statistical averaging. Curie temperatures were determined from magnetic susceptibility peaks.

\section{Results and Discussion}
\subsection{Doping-dependent evolution of isotropic exchange coupling}
\noindent
To understand how doping affects the magnetic properties of CrI$_3$, we first examine the evolution of isotropic exchange interactions. As obtained from our four-state calculations on undoped CrI$_3$, the exchange constants between first and second NN are FM, with $J_1 = 2.78\, {\rm meV}$ and $J_2 = 0.63 \, {\rm meV}$, respectively --  whereas third NN coupling $J_3 = -0.10 \, {\rm meV}$ is AFM but much weaker than these two. Our results are consistent with that of Singh \textit{et al.}~\cite{Singh2021Jun} who obtained $J_1 = 2.36\, {\rm meV}$ and $J_2 = 0.64 \, {\rm meV}$, while their $J_3$, although AFM, is of the same magnitude as $J_2$. Yet, the discrepancy between our results and that of Singh \textit{et al.} are somewhat expected as they used a different method to calculate the exchange coupling constants and have also added the Hubbard $U$ to Cr $3d$ orbitals, which we did not. Further, from the \texttt{TB2J} calculations, we obtain negligible interactions between the fourth and the more distant neighbors, as discussed in section III of the supplementary information. Therefore, we neglect the far distant exchange interactions in all the subsequent analysis.

\begin{figure}
    \centering
    \includegraphics[width=0.85\linewidth]{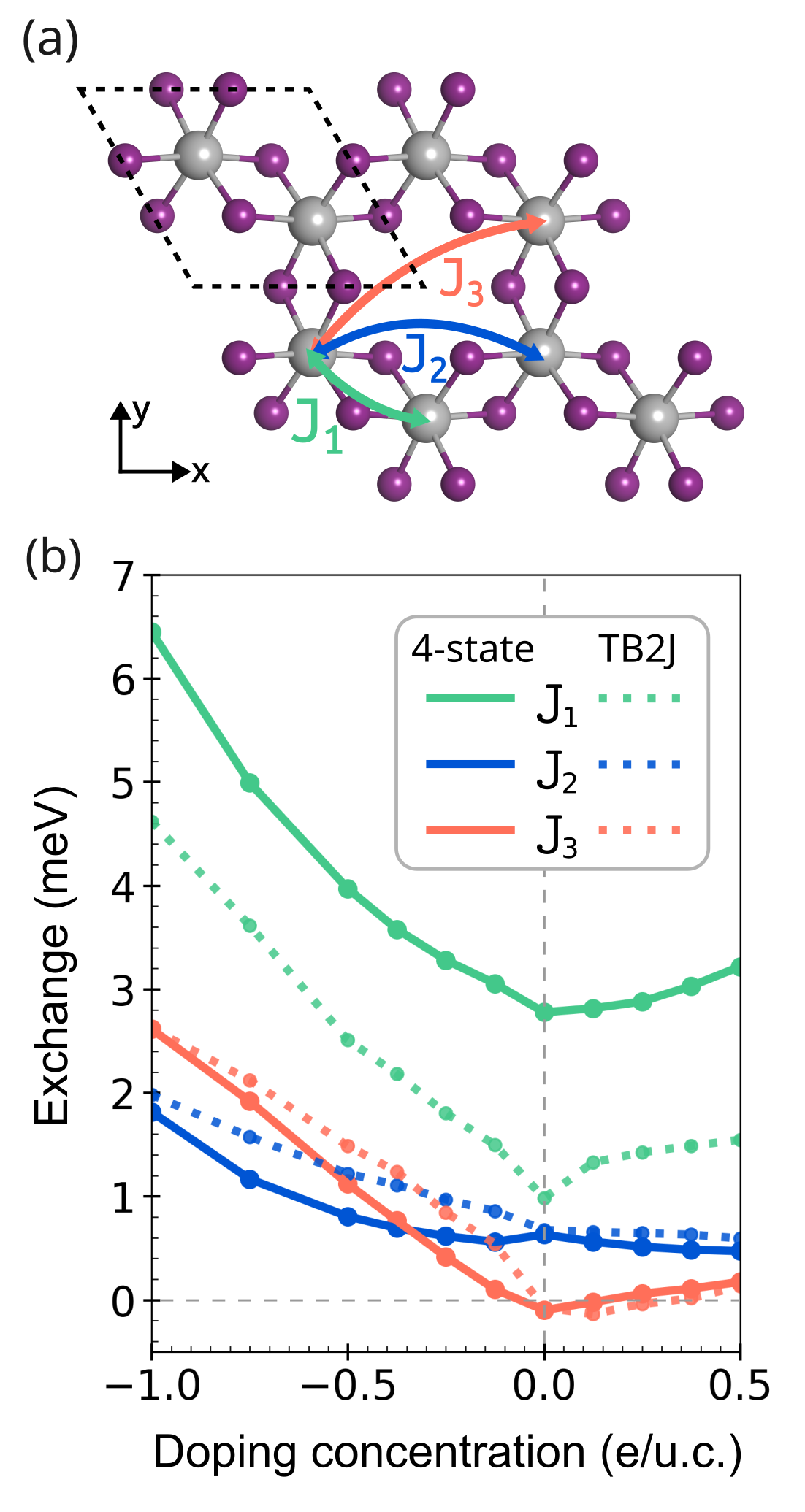}
    \caption{(a) Pairs of nearest-neighbors for which the exchange tensors are calculated in the depicted coordinate frame. (b) Doping dependency of isotropic exchanges between the first, second, and third nearest neighbors calculated by two different methodologies.}
    \label{fig:isotropic}
\end{figure}

Hole doping strongly enhances the $J_1$ coupling constant, as depicted in figure~\ref{fig:isotropic}(b), which shows that this exchange interaction increases by 43\% at the doping of 0.5 holes/u.c.
and doubles at a doping of 1 holes/u.c. In contrast, electron doping is far less efficient as $J_1$ increases by only 16\% relative to the undoped case at 0.5 electrons/u.c. Similarly to $J_1$, hole doping is highly efficient in increasing the $J_2$, whereas electron doping, surprisingly, has a negative effect on $J_2$, reducing it by 25\% at 0.5 electrons/u.c. The third NN exchange $J_3$, which is AFM for the undoped CrI$_3$, is highly sensitive to doping as it changes sign even at very small doping concentrations irrespective of the carrier type. After becoming FM, $J_3$ follows the same rate of change as $J_1$, as shown in figure~\ref{fig:isotropic}(b). 
 
Understanding the doping-induced changes of isotropic exchange coupling requires a detailed analysis of different orbital paths that govern these interactions. Although the four-state method precisely maps total energies to exchange parameters, it does not give access to the orbital decomposition of the coupling parameters. To overcome this limitation, we employ Lichtenstein formalism implemented in \texttt{TB2J}, which provides orbital-resolved exchange interactions. In comparison to $J_i$ parameters calculated with the four-state method, the values calculated using the \texttt{TB2J} method are presented with dotted lines in figure~\ref{fig:isotropic}(b). For the undoped system, \texttt{TB2J} yields $J_1 = 0.98\,\text{meV}$, significantly lower than the value obtained with the four-state method. For the doped system, while \texttt{TB2J} systematically underestimates $J_1$ compared to the four-state approach, both methods consistently predict the same qualitative trend: a rapid increase with hole doping of all exchange parameters versus a much slower increase (decrease) of $J_1$ and $J_3$ ($J_2$) under electron doping.
Despite quantitative differences between the two theoretical approaches, the strong qualitative agreement in predicting doping-induced trends in isotropic exchange parameters justifies the application of \texttt{TB2J} for orbital-resolved analysis of exchange interactions under carrier doping.

\begin{figure}[h]
    \centering
    \includegraphics[width=1.0\linewidth]{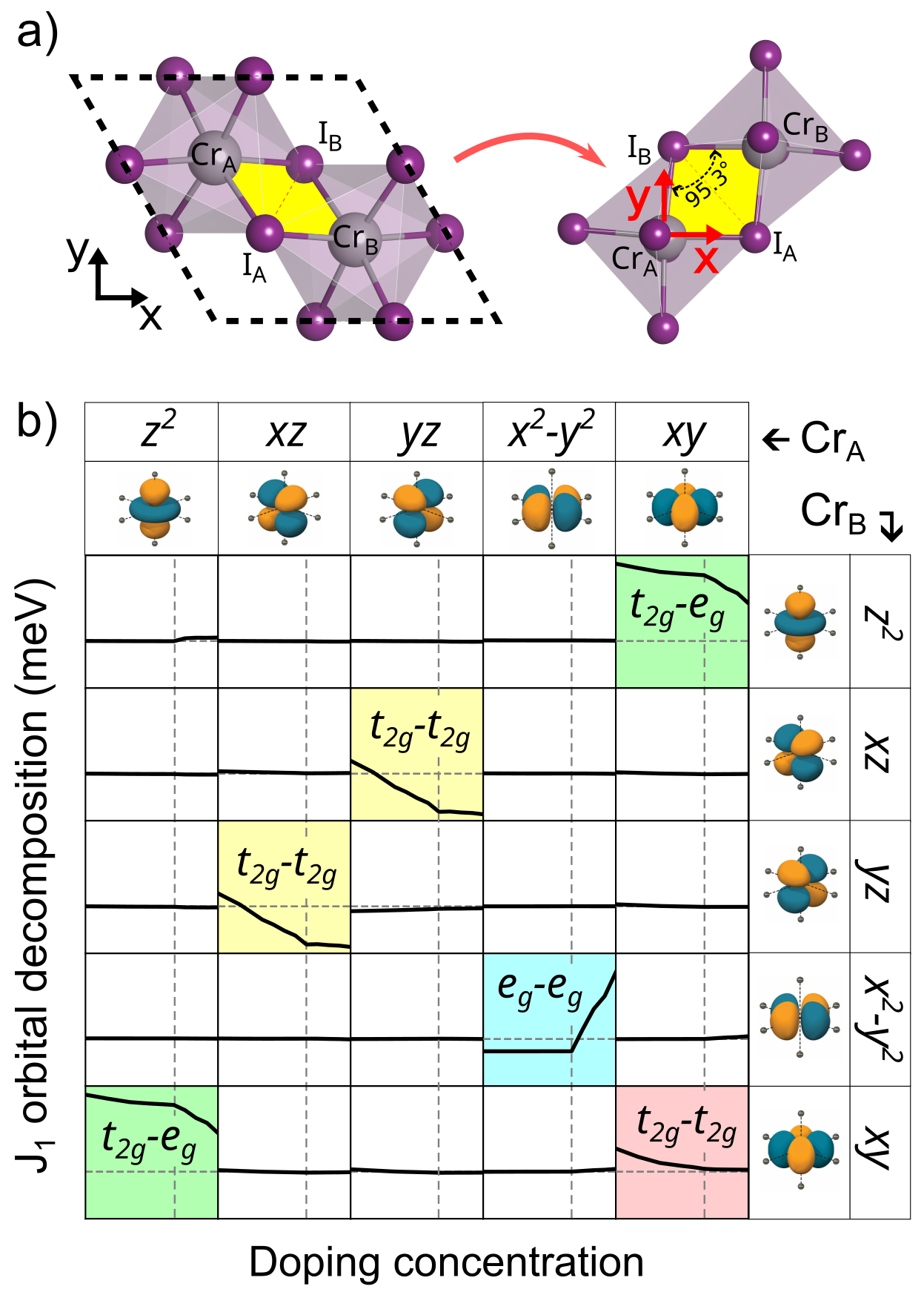}
    \caption{(a) Reorientation of the local coordinate system in the unit cell to match the direction of Cr-I chemical bonds. (b) Decomposition of the Cr-Cr superexchange into distinct $d$ orbitals across the doping range from 1 holes/u.c. to 0.5 electrons/u.c. Non-negligible interaction pairs for the electron and hole doping are highlighted with colors.}
    \label{fig:decomposition}
\end{figure}

The orbital decomposition of the exchange interaction is most naturally interpreted in a local coordinate system aligned with the Cr--I bond directions, where all superexchange pathways lie within the $xy$-plane. As shown by red arrows in figure ~\ref{fig:decomposition}(a), the axes of this coordinate system align approximately with the principal bonding directions of the CrI$_6$ octahedra. The deviation from perfect alignment arises from the Cr-I-Cr bond angles of $95.3^\circ$, which distort the octahedral symmetry. To understand trends presented in figure~\ref{fig:isotropic}(b), we focus on the first NN exchange coupling $J_1$ as it represents the dominant interaction in the system. The analysis of further-neighbor exchanges would include superexchange pathways that involve multiple intermediate ligands which makes the hopping contributions substantially more complicated to rationalize. 

Orbital decomposition of $J_1$ into contributions from different $d$-orbitals of the magnetic atoms Cr$_A$ and Cr$_B$ are presented in figure~\ref{fig:decomposition}(b). In the undoped system, most exchange pathways, like $d_{x^2-y^2} - d_{z^2}$, contribute negligibly to $J_1$ (white squares in figure~\ref{fig:decomposition}(b)). Notably, most of these exchange channels remain inert under doping, revealing that the relative orientation of $d$-orbitals on adjacent Cr atoms---a property dependent on the system's geometry and unaffected by carrier doping---plays a decisive role in establishing efficient exchange pathways. 
In total, only 6 out of 25 possible $d-d$ superexchange paths contribute to $J_1$. 

\begin{figure}
    \centering
    \includegraphics[width=0.7\linewidth]{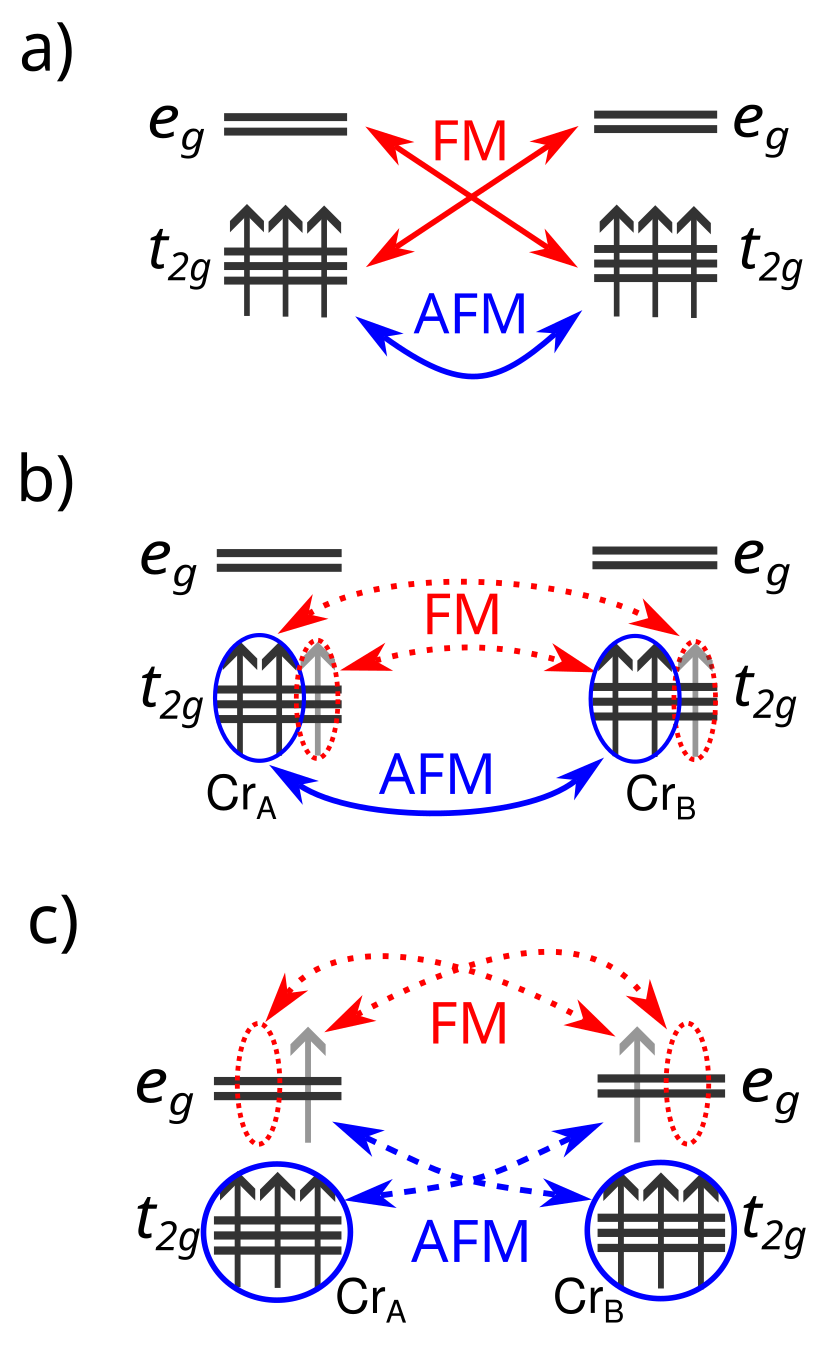}
    \caption{a) Schematic representation of competing FM and AFM interactions in undoped ${\rm CrI_3}$. b) Additional FM exchanges emerging from the interaction of occupied and partially empty $t_{2g}$ states under hole doping. c) Electron doping gives rise to new FM (AFM) exchange, represented through dotted arrows, arising from interaction of partially occupied $e_g$ with empty $e_g$ (occupied $t_{2g}$) states.}
    \label{fig:schemes}
\end{figure}

The selective contribution of specific exchange pathways to $J_1$, color-highlighted in figure~\ref{fig:decomposition}(b), can be qualitatively explained by invoking Anderson's superexchange theory~\cite{Anderson1959Jul}. According to Anderson, the interaction between two half-filled orbitals is AFM with coupling proportional to $J_{\rm AFM} \sim -t^2/U $, where $t$ is the hopping integral and $U$ is the Hubbard parameter, describing effectively the on-site Coulomb repulsion energy between electrons belonging to the same localized orbital of a magnetic atom. In undoped CrI$_3$, this kind of AFM interaction is realized by $t_{2g}-t_{2g}$ coupling, schematically represented in figure~\ref{fig:schemes}(a), with exchange constant $J_{d_{yz}-d_{xz}} = -0.89 \, {\rm meV}$ as estimated by our \texttt{TB2J} calculations.
Further, according to Anderson's theory, the interaction between the half-full and empty orbital is FM with coupling constant $J_{\rm FM} \sim t^2J_H/U^2$, where 
$J_H$ is Hund's intra-atomic exchange that favors FM spin alignment on an atom hosting the empty orbital. The corresponding FM interaction in ${\rm CrI_3}$ acts between occupied $t_{2g}$ and empty $e_g$ orbitals, illustrated by red arrows in figure~\ref{fig:schemes}(a), with a coupling constant of $J_{t_{2g}-e_g} = 1.54 \, {\rm meV}$.
Although the Cr $e_g$ orbitals are unoccupied and would nominally preclude any $e_g$--$e_g$ interaction, our \texttt{TB2J} calculations reveal weak but non-negligible AFM coupling ($J_{e_g-e_g} = -0.29 \, {\rm meV}$). This finding aligns with the results reported by \v{S}abani \textit{et al.}~\cite{Sabani2025Jun}, suggesting that higher-order processes or subtle hybridization effects may enable this unexpected exchange pathway.

\begin{figure*}
    
    \centering
    \includegraphics[width=1.0\linewidth]{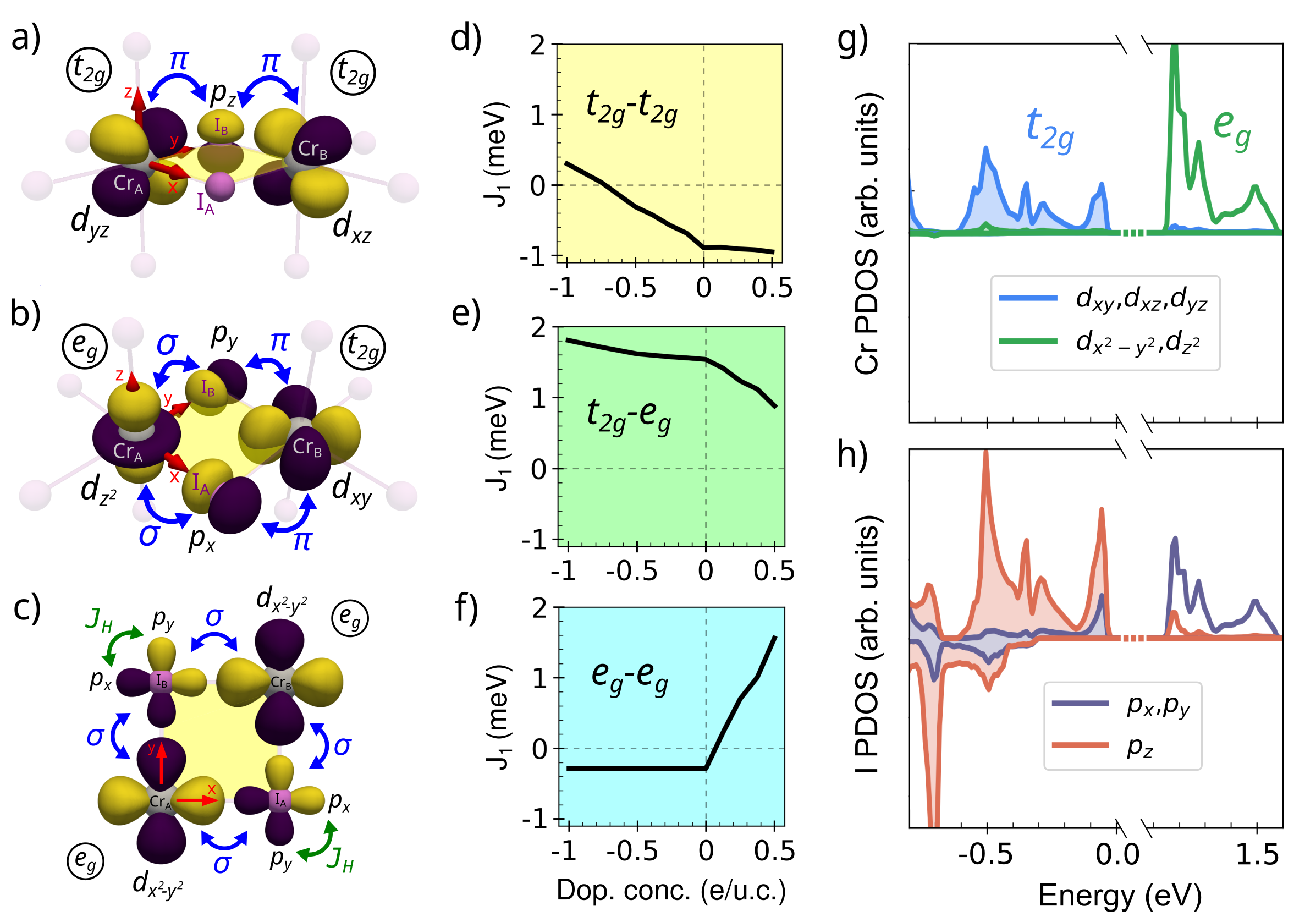}
    \caption{(a-c) Illustrated superexchange pathways between a) $t_{2g}-t_{2g}$, b) $t_{2g}-e_{g}$ and c) $e_g-e_g$ orbitals. (d-f) Corresponding orbitally-resolved exchange interactions under doping conditions. (g,h) Projected density of states (PDOS) in the rotated coordinate system (see figure~\ref{fig:decomposition}(a)) for g) Cr $d$ and h) I $p$ states.}
    \label{fig:pathways}
\end{figure*}

The superexchange pathways between the magnetic Cr atoms are mediated by I $5p$ orbitals. Concerning the $t_{2g}-t_{2g}$ superexchange between the $d_{yz}$ and $d_{xz}$ orbitals, only the $p_z$ orbital on the ${\rm I_B}$ atom overlaps with both of these $d$-orbitals, forming lateral $\pi$ bonds as illustrated in figure~\ref{fig:pathways}(a). On the other hand, the $e_g-t_{2g}$ superexchange is mediated over two distinct pathways: through the $p_x$ orbital on ${\rm I_A}$ and through the $p_y$ orbital on ${\rm I_B}$, which are both forming $\sigma$ bonds with $d_{z^2}$ on Cr$_{\rm A}$ and $\pi$ bonds with $d_{xz}$ on Cr$_{\rm B}$ atom (figure~\ref{fig:pathways}(b)). Regarding the $e_g-e_g$ interaction, $p_x$ and $p_y$ orbitals on the same I atom bridge the $d_{x^2-y^2}$ orbitals on two neighboring Cr atoms (figure~\ref{fig:pathways}(c)). At first glance, superexchange between two $d_{x^2-y^2}$ orbitals appears to be suppressed, since the electron hopping between two orthogonal orbitals $p_x$ and $p_y$ on the I atom is forbidden. However, superexchange between two $d_{x^2-y^2}$ orbitals can still occur, facilitated by on-site Hund exchange $J_H$ between unpaired spins in $p_x$ and $p_y$ orbitals~\cite{Kanamori1959Jul}. 
As we will see in the remainder of this subsection, the $e_g - e_g$ exchange path---dormant in the undoped case due to empty Cr $e_g$ orbitals---is activated by electron doping.

Building on the same framework, we proceed to analyze doping-induced changes in orbital-resolved exchange interactions. The atomic orbital--projected density of states (pDOS) plot in figures~\ref{fig:pathways}(g) and (h) demonstrates that hole and electron doping change the electron count on different atomic orbitals. The hole doping leads to partial depletion of Cr $t_{2g}$ orbitals---$d_{xy}$, $d_{xz}$, and $d_{yz}$---but it also reduces electrons count on all three I $p$ orbitals---with $p_z$ orbital being affected the most. On the other hand, electron doping populates the Cr $e_g$ orbitals---$d_{z^2}$ and $d_{x^2-y^2}$---as well as I $p_x$ and $p_y$ orbitals. 

The $t_{2g}-t_{2g}$ interaction between $d_{yz}$ and $d_{xz}$ orbitals, being AFM in undoped CrI$_3$, undergoes a progressive change of its character under hole doping, culminating in a full AFM-to-FM reversal at a doping concentration of $\sim 0.75$ holes/u.c., as illustrated in figure~\ref{fig:pathways}(d). 
This FM enhancement arises from exchange interactions between a partially vacant $t_{2g}$ orbital on one Cr atom and an occupied $t_{2g}$ orbital on a neighboring Cr atom, as illustrated schematically in figure~\ref{fig:schemes}(b). The strength of this interaction grows progressively with increasing hole doping concentration, as occupancy of the partially vacant orbital decreases. In contrast to hole doping, electron doping leaves the Cr $t_{2g}$ orbitals unaffected, resulting in negligible changes to the $J_{t_{2g}-t_{2g}}$ coupling constant.

The influence of doping on the $t_{2g}-e_g$ exchange interaction further highlights the asymmetric response of ${\rm CrI_3}$ to doping with different types of charge carriers. Hole doping has a minor impact on the $t_{2g}-e_g$ exchange (figure~\ref{fig:pathways}(e)) given that the mediating $p_x$ and $p_y$ states are situated predominantly in the conduction region and are thus not affected by hole doping (figure~\ref{fig:pathways}(h)). In contrast, electron doping strongly suppresses the FM interaction, reducing its value by 43\% at the doping concentration of 0.5 electrons/u.c. This suppression arises from an additional AFM interaction between electrons in $t_{2g}$ orbitals and those in partially filled $e_g$ orbitals, as represented with dotted blue lines in figure~\ref{fig:schemes}(c). The $t_{2g}-e_g$ channel becomes active for electron doping because $p_x$ and $p_y$ orbitals on I atoms are being populated.

The third exchange type $e_g-e_g$ is naturally not affected by hole doping as the $e_g$ orbitals are completely empty in the undoped system. On the contrary, electron doping dramatically increases the FM exchange in the $e_g-e_g$ channel, as evidenced in figure~\ref{fig:pathways}(f). This is because the electron in the $e_g$ orbital on the one Cr atom ferromagnetically couples to the empty $e_g$ orbital on the other Cr atom, as schematically presented in figure~\ref{fig:schemes}(c). The $e_g-e_g$ exchange exhibits drastic enhancement compared to other exchange channels, as electron doping directly affects all the key orbitals involved in this interaction: the Cr $d_{x^2-y^2}$ orbitals, along with I $p_x$/$p_y$ orbitals which mediate this superexchange interactions (see figure~\ref{fig:pathways}(c) for the interaction pathway and figures~\ref{fig:pathways}(g) and (h) for the pDOS). 

The last contribution to $J_1$ is the weakest of the four. It is purely FM and activates exclusively under hole doping, growing monotonically with doping concentration (red square in figure~\ref{fig:decomposition}(b)). This $d_{xy}-d_{xy}$ superexchange, mediated by $\pi$-bonded $p_x/p_y$ orbitals (similar to the $e_g-e_g$ pathway in figure~\ref{fig:pathways}(c)), remains inactive in undoped CrI$_3$ and cannot be triggered by electron doping due to complete occupation of the Cr $d_{xy}$ orbitals, which suppresses the necessary hopping processes. Its ferromagnetic character originates from $t_{2g}-t_{2g}$ exchange denoted by red arrows in figure~\ref{fig:schemes}(b). 

In summary, the overall enhancement of nearest-neighbor FM exchange under hole doping stems primarily from the progressive suppression of the AFM $t_{2g}-t_{2g}$ interaction channels. In contrast, in an electron-doped system, the strengthening of a single FM $e_g-e_g$ interaction is largely offset by the weakening of two FM $t_{2g}-e_g$ couplings, leading to negligible net changes in $J_1$.

\subsection{SOC--induced magnetic properties under doping}
\noindent
SOC is the source of magnetic anisotropy that plays a major role in stabilizing long-range magnetic order in 2D magnets~\cite{Mermin1966Nov}. In this subsection, we systematically investigate how charge carrier doping modifies MAE and its constituents including SIA, symmetric anisotropic exchange, and DMI. Finally, we analyze how doping affects the $T_{\rm c}$ which depends both on the isotropic exchange, discussed in the previous subsection, and on these SOC-dependent properties. 
\begin{figure}
    \centering
    \includegraphics[width=1.0\linewidth]{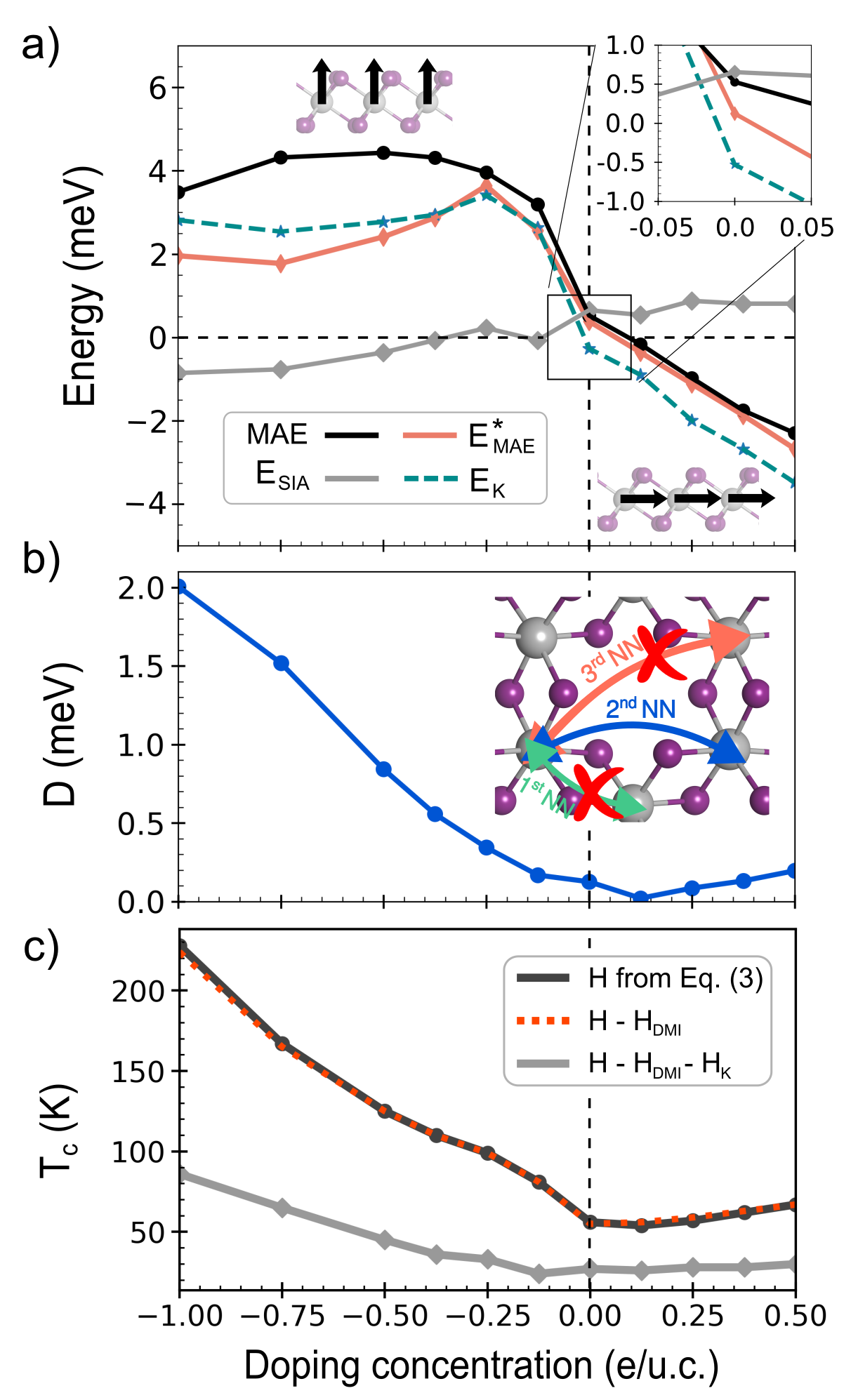}
    \caption{a) MAE and SIA show opposite signs in both electron and hole doping regions, suggesting different easy-axis orientations. Modeled MAE ($E^*_{\rm MAE}$) includes symmetric anisotropic exchange $E_{\rm K}$, solving the discrepancy in the entire doping range between the calculated MAE and SIA. b) Magnitude of the vector ${\bf D}$ at different doping concentrations; the inset indicates the interaction is prohibited among the first and third NN. c) Doping-dependent evolution of Curie temperatures computed using the Hamiltonians specified in the legend.}
    \label{fig:anisotropy}
\end{figure}

SIA refers to the energy associated with the orientation of an individual magnetic ion's spin relative to specific directions in the crystal lattice. It arises from the SOC acting at the atomic level, in combination with the symmetry of the local crystal field surrounding the magnetic ion. 
In the following, we express SIA in terms of energy $E_{\rm SIA} = A_{zz}S^2 = \frac{9}{4}A_{zz}$, rather than the coupling constant $A_{zz}$.
In undoped CrI$_3$, we obtained the positive value of $E_{\rm SIA} =  0.65 \, \text{meV}$, which infers an out-of-plane easy axis for Cr spins, in agreement with previous studies~\cite{Xu2018Nov,Singh2021Jun}. 
Electron doping barely affects SIA, as $E_{\rm SIA}$ remains positive across all inspected concentrations of electron dopants, ranging from $0.54$ to $0.88\,\mathrm{meV}$ (figure~\ref{fig:anisotropy}(a)). Therefore, under electron doping, iodine's crystal field consistently favors the out-of-plane orientation of the Cr spin.
In sharp contrast to electron doping, hole doping reverses the sign of $E_{\rm SIA}$, favoring in-plane spin alignment. For strong hole doping the in-plane SIA becomes even more pronounced as $E_{\rm SIA}$ reaches the value of $-0.86 \, \text{meV}$ at 1 holes/u.c.
The pronounced SIA modification under hole doping likely stems from changes in I $5p$ orbital occupation, which perturbs the crystal field at Cr sites. In contrast, electron doping populates the (initially empty) Cr $e_g$ orbitals without significantly affecting the iodine crystal field, leaving the SIA essentially unchanged.

However, it is not SIA but MAE that ultimately determines the preferred direction of magnetic moments. MAE is defined as the energy difference between states with in-plane and out-of-plane magnetization, $E_{\rm MAE} = E_{\parallel}- E_{\bot}$. 
In undoped CrI$_3$ we obtain a positive $E_{\rm MAE}$ of $0.53 \, {\rm meV}$ per Cr atom, which aligns with the $E_{\rm SIA}$ of $0.65 \, {\rm meV}$, confirming an out-of-plane easy axis as shown in figure~\ref{fig:anisotropy}(a). This is in agreement with previous studies which reported the MAE of $0.65\,{\rm meV}$~\cite{Lado2017Jun}. However, hole doping dramatically alters this picture inducing a significant enhancement of the MAE while simultaneously reducing the SIA (figure~\ref{fig:anisotropy}(a)). This apparent contradiction suggests competing mechanisms govern the doping-dependent anisotropy evolution. 
After reaching a maximum of $4.43 \,\mathrm{meV}$ at 0.5 holes/u.c., the MAE shows a monotonic decrease with additional hole doping. In stark contrast, electron doping maintains the SIA nearly unchanged but dramatically suppresses the MAE. At doping levels above 0.1 electrons/u.c. this reduction becomes so substantial that the system transitions to in-plane ferromagnetism. As electron doping increases, the MAE exhibits a linear decrease (with $E_{\text{MAE}}$ growing progressively more negative), further stabilizing the in-plane magnetic orientation.

To understand opposite trends between the behavior of SIA and MAE under doping, we estimate MAE from the spin Hamiltonian exposed in equation~(\ref{eq:H_final}) and compare it to DFT-calculated MAE. This approach enables unambiguous separation of the SIA and anisotropic exchange contributions to the MAE. To this end we consider two distinct spin configurations, with all the spins in the system being either ${\bf S} = (0,3/2,0)$ or ${\bf S} = (0,0,3/2)$. Inserting these spins in the Hamiltonian from equation~(\ref{eq:H_final}) and subtracting the two obtained energies yields 
\begin{equation}
\label{eq:MAE_model}
E_{\rm MAE}^* = \mathcal{H}_{\parallel} - \mathcal{H}_{\perp} = \mathcal{H}_{\rm K,\parallel}-\mathcal{H}_{\rm K,\perp}+E_{\rm SIA},
\end{equation}
where we introduce an auxiliary function quantifying the symmetric anisotropic exchange of a spin ${\bf S}_i$ with its neighbors, 
\begin{equation}
\label{eq:K}
\begin{aligned}
    \mathcal{H}_{{\rm K},i} = &- \sum_{j=1,2,3} {\bf S}_i\mathcal{K}_{1,ij}{\bf S}_j 
    - \sum_{j=1,\ldots,6} {\bf S}_i\mathcal{K}_{2,ij}{\bf S}_j \\
    &-  \sum_{j=1,2,3} {\bf S}_i\mathcal{K}_{3,ij}{\bf S}_j.
\end{aligned}
\end{equation}
In the last equation the exchange tensors $\mathcal{K}_{n,ij}$ ($n=1,2,3$ denote neighbors) depend on the bond directions between Cr sites $i$ and $j$ and are computed with the four-state method. Note that the isotropic exchange and the DMI do not contribute to $E_{\rm MAE}^*$: the isotropic exchange contributions are the same in the two spin configurations and thus cancel exactly, while the DMI vanishes as all the spins are parallel in both considered spin configurations. 

Figure~\ref{fig:anisotropy}(a) compares our DFT-calculated $E_{\rm MAE}$ with the modeled $E^*_{\rm MAE}$, demonstrating excellent agreement in easy-axis orientation across all doping levels. Notably, the anisotropic exchange contribution dominates the MAE, particularly under hole doping, surpassing the SIA by a significant margin. While Xu \textit{et al.}~\cite{Xu2018Nov} first highlighted the crucial role of anisotropic exchange in undoped CrI$_3$---showing how its interplay with SIA governs spin orientation---our work reveals its enhanced significance in doped systems, where it accounts for nearly the entire MAE. 

We note that for hole doping concentrations exceeding 0.25 holes/u.c., we observe a systematic deviation where the modeled MAE underestimates the DFT-calculated values. This discrepancy suggests that in strongly hole-doped CrI$_3$ new anisotropic contributions emerge, that are not accounted for in the spin Hamiltonian exposed in equation~(\ref{eq:H_final}). While discrepancies arise under strong hole doping, the direct calculations and model predictions for MAE maintain excellent agreement across experimentally realistic doping levels. We note that the achievable doping range in experiments (e.g., $0.11$ holes/u.c. to $0.11$ electrons/u.c. in Ref.~\cite{Jiang2018Jul}) falls within this regime of quantitative agreement.

As mentioned in subsection~\ref{sseq:hamiltonian}, Moriya's symmetry rules forbid DMI between first- and third-neighbor Cr sites in CrI$_3$ while permitting it between second neighbors, as schematically illustrated in the inset of figure~\ref{fig:anisotropy}(b). As we keep the atomic positions fixed, the DMI constraints remain identical for both undoped and doped systems. For each second-nearest-neighbor Cr pair, Moriya's rules constrain the vector $\mathbf{D}$ to lie within the plane orthogonal to the material's plane and containing the bond direction of the pair.
The magnitude of the vector ${\bf D}$ for the undoped system is only $D = 0.12 \, {\rm meV}$ -- which is the reason why this anisotropic interaction is neglected in most computational studies on CrI$_3$. We note that in doped systems, as obtained explicitly from our four-state calculations, the magnitude and the direction of vector ${\bf D}$ vary but the vector itself (as expected) stays within the plane determined by Moriya's rules. While electron doping has minimal influence on the strength of the DMI, with $D$ not exceeding $0.20 \, {\rm meV}$ even at high electron concentrations, hole doping significantly enhances it, increasing $D$ to $2.01 \, {\rm meV}$ at a doping concentration of 1 holes/u.c., as presented in figure~\ref{fig:anisotropy}(a). To put it in perspective of exchange interactions, strong hole-doping increases the DMI to the order of magnitude of isotropic exchange interactions with ratios $D/J_1 \approx 0.3$ and $D/J_2 \approx 1.1$.

The enhancement of DMI between the second NN due to doping is in fact an expected outcome as in conductive environment the presence of itinerant electrons allows DMI (and other exchange interactions) to be mediated over longer distances. On the contrary, in semiconductors (like in undoped CrI$_3$) the lack of itinerant carriers means that magnetic exchange (and DMI) is primarily mediated by superexchange or similar localized mechanisms. 
The superior efficiency of hole doping in increasing DMI can be explained by the fact that hole doping primarily affects the delocalized I $p$ orbitals and related highly dispersive electronic bands, whereas electron doping predominantly influences the much-more localized Cr $e_{g}$ orbitals and their flat-like electronic bands.
Moreover, a sharp increase of DMI between the second NN fully justifies the use of a tensor formalism for the description of distant-neighbor exchange in doped 2D magnetic semiconductors.

Finally, having discussed isotropic and anisotropic exchange interactions as well as magnetic anisotropy, we examine how doping affects the Curie temperature, where all these contributions play a role. The Curie temperature is calculated across the entire doping range using Monte Carlo simulations with the spin Hamiltonian defined in equation~(\ref{eq:H_final}). For undoped ${\rm CrI_3}$ we obtain $T_c = 56\,{\rm K}$, which is a satisfactory result compared to the experimental value of $45\,{\rm K}$. Hole doping leads to a significant increase in both exchange interactions and MAE, resulting in a substantial enhancement of $T_c$, which reaches $228\,{\rm K}$ at 1 holes/u.c.
-- more than four times the value in the undoped system (figure~\ref{fig:anisotropy}(c)). In contrast, electron doping has a much weaker effect: $T_c$ remains below $67\,{\rm K}$, reflecting only modest changes in isotropic interactions.

To assess whether a substantial increase in DMI under hole doping could influence the Curie temperature, we recalculate $T_c$ by excluding the DMI from the Hamiltonian in equation~(\ref{eq:H_final}). The results in figure~\ref{fig:anisotropy}(c) show negligible differences, confirming that the strong isotropic exchange among the nearest neighbors dominates over the longer-range DMI. 
Next, we analyze how the Curie temperature reflects the competition of the other two anisotropic contributions, SIA and symmetric anisotropic exchange. Monte Carlo simulations, using only isotropic exchanges and SIA from equation~(\ref{eq:H_final}), show a significant $T_c$ drop to $27\, {\rm K}$ for the undoped case. In this simplified $J + A$ model, the $T_c$ is saturated for electron doping, whereas for hole doping it increases from $27\, {\rm K}$ to $86\,{\rm K}$ (gray line in figure~\ref{fig:anisotropy}(c)). This increase is mainly due to the enhancement of isotropic exchanges since the SIA under hole doping does not exceed the magnitude it reaches for electron doping. Therefore, disabling the symmetric anisotropic exchange would reduce $T_c$ by a factor of several, demonstrating that this anisotropic interaction is indeed responsible for the dramatic Curie temperature enhancement in strongly hole-doped CrI$_3$ monolayer.

\section{Conclusion}
\noindent
In conclusion, our comprehensive theoretical study reveals a striking electron-hole doping asymmetry of magnetic properties of monolayer CrI$_3$, originating from distinct orbital-resolved exchange mechanisms. Crucially, hole doping induces a much higher degree of electronic itinerancy compared to electron doping, which retains a more localized, undoped-like character of CrI$_3$. We demonstrate that the $t_{2g}-t_{2g}$ exchange interaction channel, initially strongly AFM in undoped CrI$_3$, weakens progressively with increasing hole doping concentration, eventually transitioning to FM character. This reveals that the enhancement of FM exchange in hole-doped CrI$_3$ originates fundamentally from the suppression of AFM interactions. In contrast, electron doping strengthens the $e_g-e_g$ FM exchange while simultaneously weakening $t_{2g}-e_g$ FM exchange, resulting in minimal net impact on exchange coupling constants. Beyond strengthening the isotropic exchange interaction, hole doping dramatically amplifies the symmetric anisotropic exchange, hence substantially increasing MAE and stabilizing out-of-plane magnetization. The cooperative enhancement of both isotropic and anisotropic exchange interactions under strong hole doping leads to a dramatic fourfold increase in $T_c$. Although we focus on monolayer CrI$_3$ as a paradigmatic 2D ferromagnetic semiconductor, our findings should generalize to similar vdW magnetic materials possessing dispersive valence bands and relatively flat conduction bands. More broadly, this work establishes strategic doping as a powerful approach for tailoring magnetic properties and anisotropy in atomically thin vdW systems.

\section*{Acknowledgments}

M.O., \v{Z}.\v{S}., and S.S. acknowledge financial support from the Vin\v{c}a Institute, provided by the Ministry of Science, Technological Development, and Innovation of the Republic of Serbia.
B.N.Š. acknowledges support from the Ministry of Education, Science, and Innovation of Montenegro. 
S.P. acknowledges funding through the Next-Generation EU programme PRIN-2022 project “SORBET: Spin-ORBit Effects in Two-dimensional magnets” (IT MIUR grant no. 2022ZY8HJY). 
M.O., S.S., and S.P. acknowledge additional funding from the Ministry of Foreign Affairs of Italy and the Ministry of Science, Technological Development, and Innovation of Serbia through the bilateral project ``Van der Waals Heterostructures for Altermagnetic Spintronics'', which is realized under the executive programme for scientific and technological cooperation between the two countries. 
Computational resources and support were provided by CINECA under the ISCRA initiative, specifically through the project ISCRA-B HP10BA00W3.


\newpage
\onecolumngrid
\section*{References}
\twocolumngrid



\newpage

\renewcommand{\thesection}{\Roman{section}}
\renewcommand{\theequation}{S\arabic{equation}}
\renewcommand{\thefigure}{S\arabic{figure}} 
\renewcommand{\thetable}{S\arabic{table}}

\onecolumngrid
\section*{Supplementary information} 

\setcounter{figure}{0}
\setcounter{table}{0} 
\setcounter{section}{0} 
\setcounter{equation}{0} 

The supplementary information is organized into five sections. Section I presents a test ensuring that excess charges do not leak into the vacuum region upon doping for the range of concentrations adopted in our work. Section II describes the transformation of the unit cell to obtain an orthogonal one required for Monte Carlo simulations. Section III validates the Wannierization procedure by comparing the electronic band structures. Section IV demonstrates the negligible contribution of isotropic exchange interactions beyond the third NN. Finally, Section V outlines the energy mapping procedure and lists all magnetic interaction parameters obtained in this work.

\section{Localization of excess charges}

\begin{figure}[ht]
    \centering
    \includegraphics[width=0.7\linewidth]{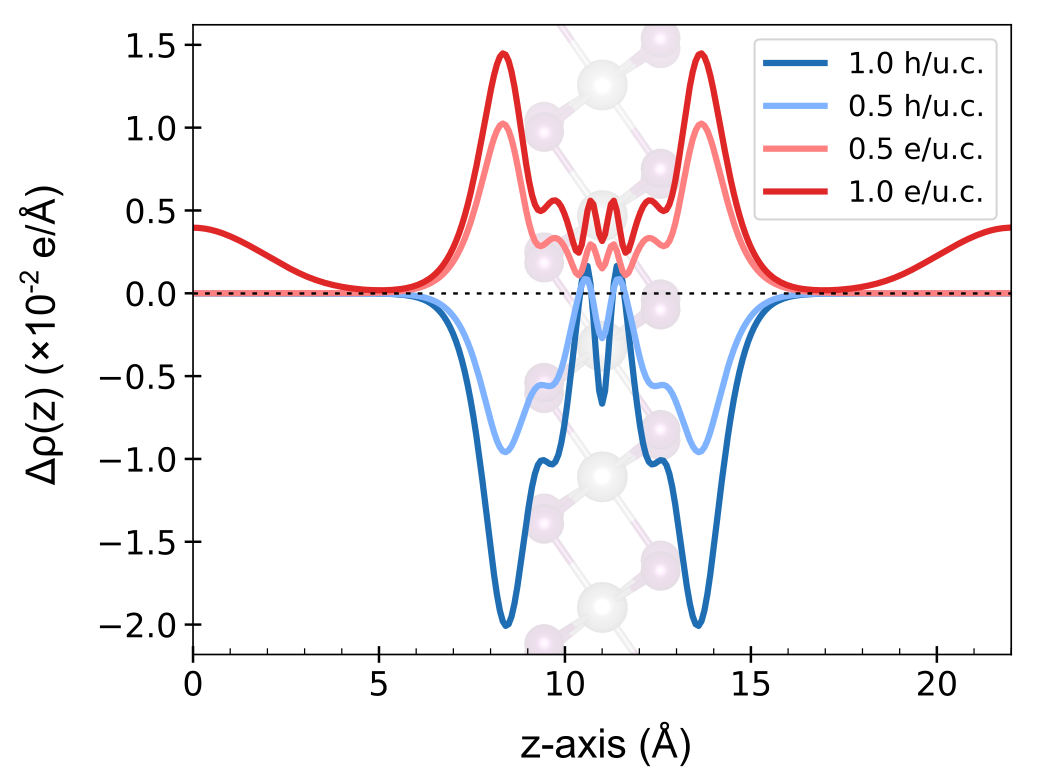}
    \caption{Excess charge density along the $z$-axis for different doping concentrations. For high electron-doping concentrations excess electrons leak into the vacuum region (red curve).}
    \label{fig:icd}
\end{figure}

Standard DFT treatment of uniform doping can introduce artificial charge distributions and potentially unreliable results. To validate our findings, we performed additional tests to verify the spatial localization of doped charges. Specifically, we computed the \textit{excess charge density} $\Delta \rho ({\bf r})$ by subtracting the undoped CrI$_3$ charge density from the doped system's charge density. Subsequently, $\Delta \rho ({\bf r})$ is averaged in the crystal plane ($xy$-plane) and its $z$-dependence, $\Delta \rho(z)$, is presented in figure~\ref{fig:icd} for various doping concentrations. These induced charges, analogous to the surface charge redistribution occurring in 2D materials upon adatom adsorption, remain localized near the atomic sites for doping levels of 1 holes/u.c., 0.5 holes/u.c., and 0.5 electrons/u.c. At high electron doping concentrations (specifically 1 electrons/u.c.), we observe artificial charge leakage into the vacuum region between periodic images. This artifact persists even when increasing the vacuum spacing from $22\,\text{\AA}$ to $30\,\text{\AA}$, revealing it to be an intrinsic limitation of the DFT doping implementation. Consequently, we restrict our analysis to the doping range between 1 holes/u.c. and 0.5 electrons/u.c., where DFT yields physically reliable results.

We also investigate the effects of vacuum spacing on the exchange interactions for several doping concentrations. We tested the vacuum thickness to assess its effects on $\mathcal{J}_1$ matrix, which represents the dominant exchange contribution in the system, as discussed in section IIIA of the main text. As shown in table~\ref{tab:J1_test}, an $18\,{\rm \AA}$ vacuum is already sufficient for obtaining reliable diagonal matrix elements of the $\mathcal{J}_1$ matrix. However, we note that off-diagonal matrix elements, representing anisotropic contributions to the exchange interaction, are more sensitive to vacuum thickness. In fact, increasing the vacuum from $18\,{\rm \AA}$ to $22\,{\rm \AA}$ results in a change of $0.1\,{\rm meV}$ in the off-diagonal elements $J_{yz}$ and $J_{zy}$ for the undoped system, which represents roughly 25\% of their value. Further increase of the vacuum separation to $26\, {\rm \AA}$ does not influence any parameters across all doping concentrations. We note that only a large vacuum of $30\,{\rm \AA}$ causes the abrupt changes in the parameters at the limiting doping concentrations of 0.5 electrons/u.c., after which the excess charges leak into the vacuum. Based on these results, we adopted $22\,{\rm \AA}$ of vacuum as sufficient spacing to avoid interaction between periodic images, both for undoped and doped systems.

\begin{table}[h]
    \caption{$J_1$ matrix elements for different vacuum spacing along z-axis and doping concentrations}
    \centering
    \begin{tabular}{|c|c|rrrrrrrrr|}
    \hline
doping ($e/u.c.$) & vacuum ($\rm \AA$) & $J_1^{xx}$ & $J_1^{yy}$ & $J_1^{zz}$ & $J_1^{xy}$ & $J_1^{yx}$ & $J_1^{xz}$ & $J_1^{zx}$ & $J_1^{yz}$ & $J_1^{zy}$\\
\hline
\multirow{4}{*}{-0.5} 
    & 18 &3.87 & 3.48 & 4.50 & -0.36 & -0.36 & -0.23 & -0.23 & -0.39 & -0.39\\
    & 22 &3.89 & 3.50 & 4.52 & -0.36 & -0.36 & -0.23 & -0.23 & -0.39 & -0.39\\
    & 26 &3.90 & 3.51 & 4.54 & -0.36 & -0.36 & -0.23 & -0.23 & -0.39 & -0.39\\
    & 30 &3.91 & 3.52 & 4.54 & -0.36 & -0.36 & -0.23 & -0.23 & -0.39 & -0.39\\ \hline
\multirow{4}{*}{0.0} 
    & 18 &2.96 & 2.55 & 2.78 & -0.37 & -0.37 & -0.22 & -0.22 & -0.27 & -0.27\\
    & 22 &2.97 & 2.54 & 2.82 & -0.36 & -0.37 & -0.21 & -0.22 & -0.38 & -0.38\\
    & 26 &2.98 & 2.55 & 2.76 & -0.36 & -0.36 & -0.21 & -0.21 & -0.38 & -0.38\\
    & 30 &2.96 & 2.55 & 2.93 & -0.38 & -0.38 & -0.24 & -0.24 & -0.41 & -0.41\\ \hline
\multirow{4}{*}{0.5} 
    & 18 &3.08 & 3.74 & 2.78 & 0.58 & 0.58 & 0.30 & 0.30 & 0.56 & 0.56\\
    & 22 &3.10 & 3.76 & 2.80 & 0.58 & 0.58 & 0.29 & 0.29 & 0.56 & 0.56\\
    & 26 &3.12 & 3.76 & 2.81 & 0.57 & 0.57 & 0.29 & 0.29 & 0.55 & 0.55\\
    & 30 &3.02 & 3.53 & 2.68 & 0.63 & 0.63 & 0.20 & 0.20 & 0.46 & 0.46\\
    \hline
    \end{tabular}
    \label{tab:J1_test}
\end{table}

\newpage
\section{Unit cell transformation for Monte Carlo simulations}
\begin{figure}[h]
    \centering
    \includegraphics[width=0.6\linewidth]{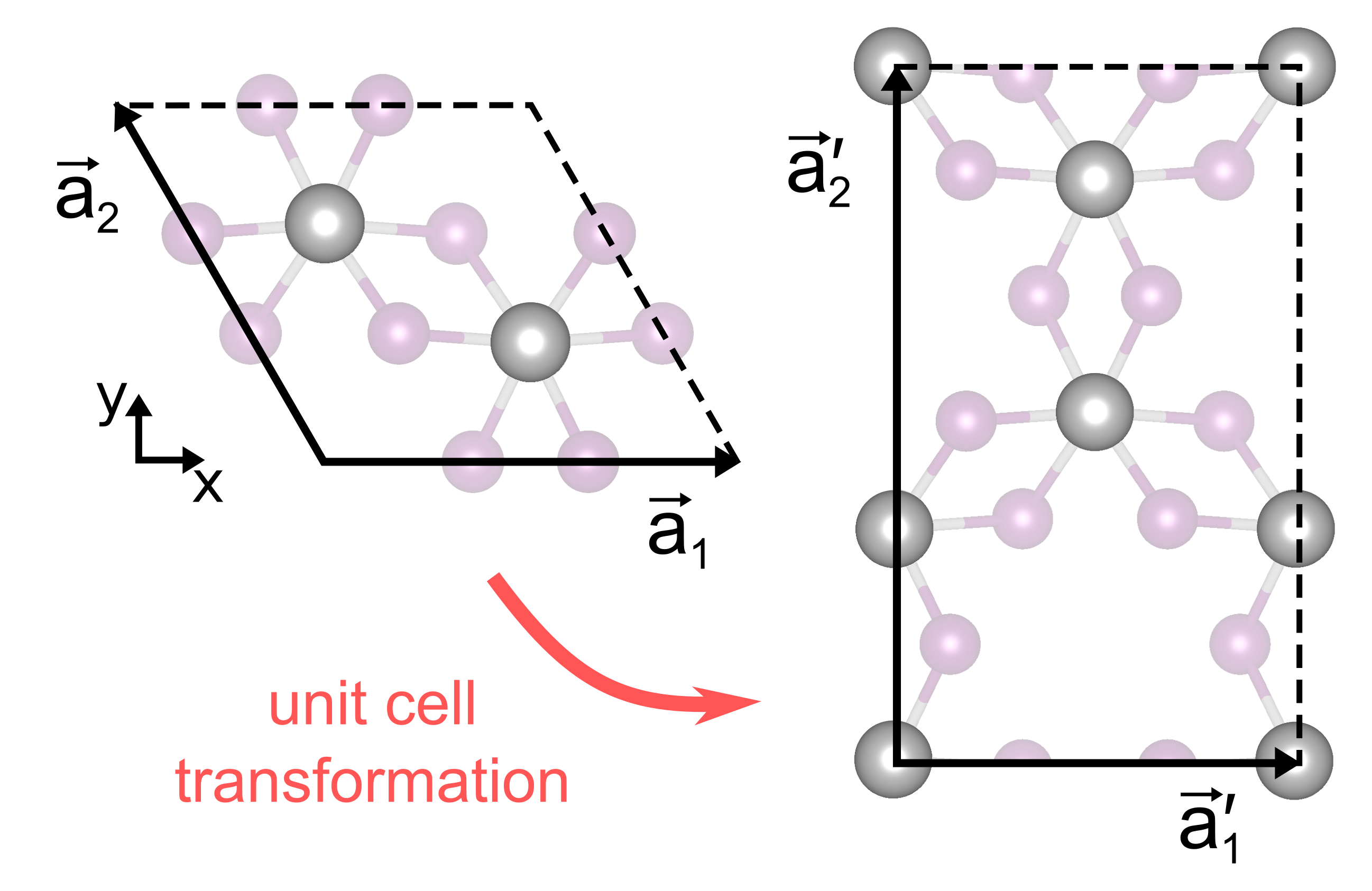}
    \caption{Hexagonal unit cell of ${\rm CrI_3}$ (left) is transformed into the rectangular unit cell (right).}
    \label{fig:uc_transformation}
\end{figure}

To assess the magnetic properties of the finite temperature spin system, we performed Monte Carlo simulations using the spin Hamiltonian from equation (3) of the main text. As the VAMPIRE software package, used for Monte Carlo simulations, requires as an input the unit cell with orthogonal lattice vectors, we transformed the original hexagonal unit cell of CrI$_3$ to a rectangular unit cell (figure~\ref{fig:uc_transformation}). Due to the honeycomb structure of the magnetic sublattice, the unit cell must be doubled and one of the lattice vectors rotated to obtain an orthogonal geometry. The matrix given by
\begin{equation}
    M =
\begin{pmatrix}
1 & 1/\sqrt{3} &0 \\
0 & 2 & 0 \\
0 &0 & 1
\end{pmatrix},
\end{equation}
transforms the primitive lattice vectors ${\bf a}_1 = (a,0,0)$ and ${\bf a}_1 = (-a/2,a\sqrt{3}/2,0)$ to orthogonal lattice vectors ${\bf a}_1' = (a,0,0)$ and ${\bf a}_2' = (0,a\sqrt{3},0)$, where the $a$ is a lattice constant of the original cell.
After applying the transformation, the origin of the unit cell is translated by $(0,a/\sqrt{3},0)$. The resulting orthogonal cell is shown on the right in figure~\ref{fig:uc_transformation}.

\section{Wannier interpolation}
\begin{figure}[h]
    \centering
    \includegraphics[width=0.6\linewidth]{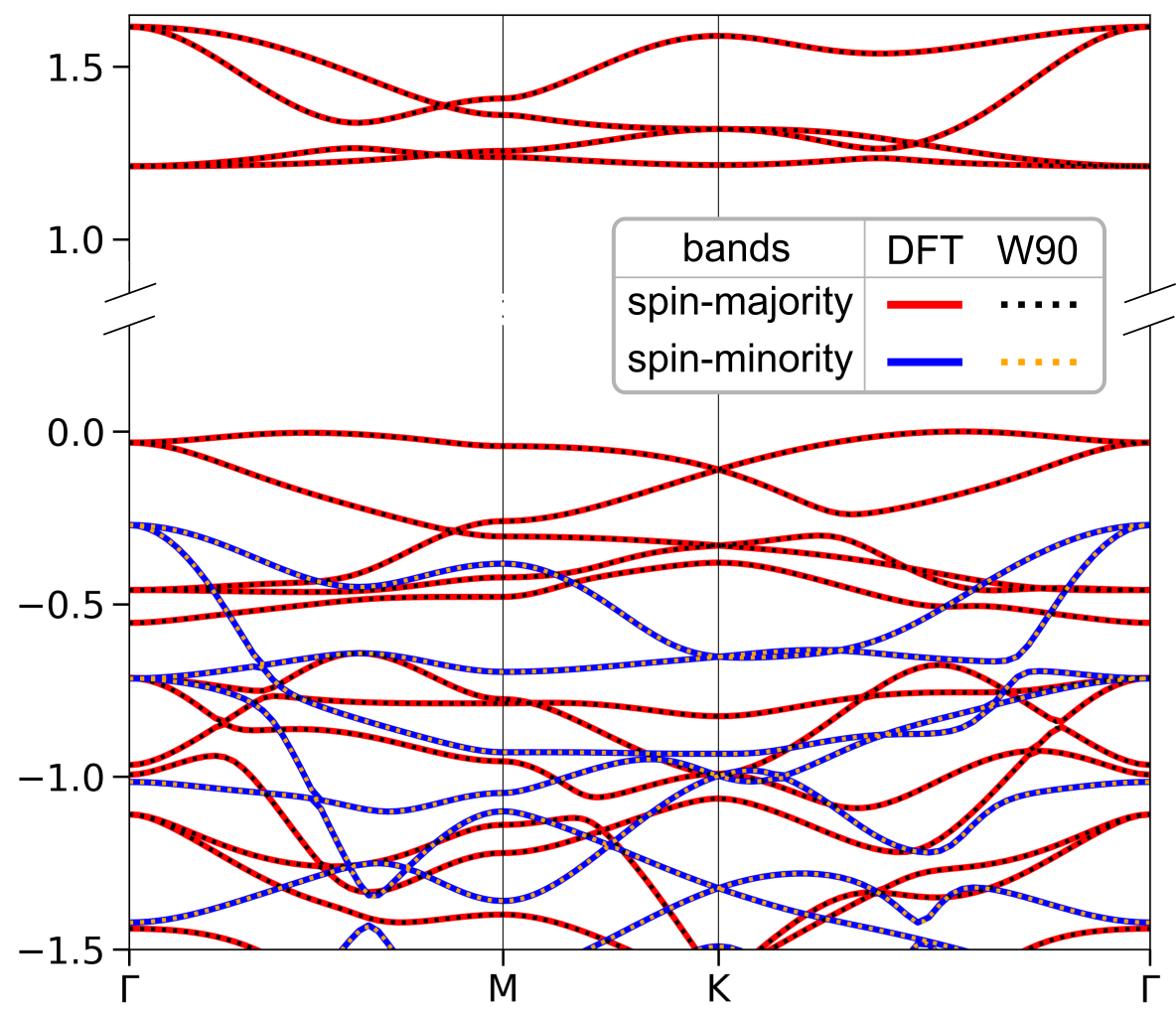}
    \caption{Band structure obtained from Wannier90 tight binding Hamiltonian is fitted to DFT band structure of ${\rm CrI_3}$.}
    \label{fig:wannier_bands}
\end{figure}

The \texttt{TB2J} results presented in section IIIA of the manuscript are dependent on the accuracy of the Wannierization procedure. A common indicator for reliable Wannierization is agreement between DFT bands and those obtained from the tight-binding Hamiltonian constructed using Wannier90. In our case, the Wannier90 bands (shown as dotted lines in figure~\ref{fig:wannier_bands}) reproduce both the spin-minority and spin-majority bands with excellent agreement. This validates the Wannierization procedure and provides confidence in the accuracy of the exchange parameters obtained from \texttt{TB2J}.

\section{Contribution from distant neighbors to the exchange interactions}

\begin{figure}[h]
    \centering
    \includegraphics[width=0.95\linewidth]{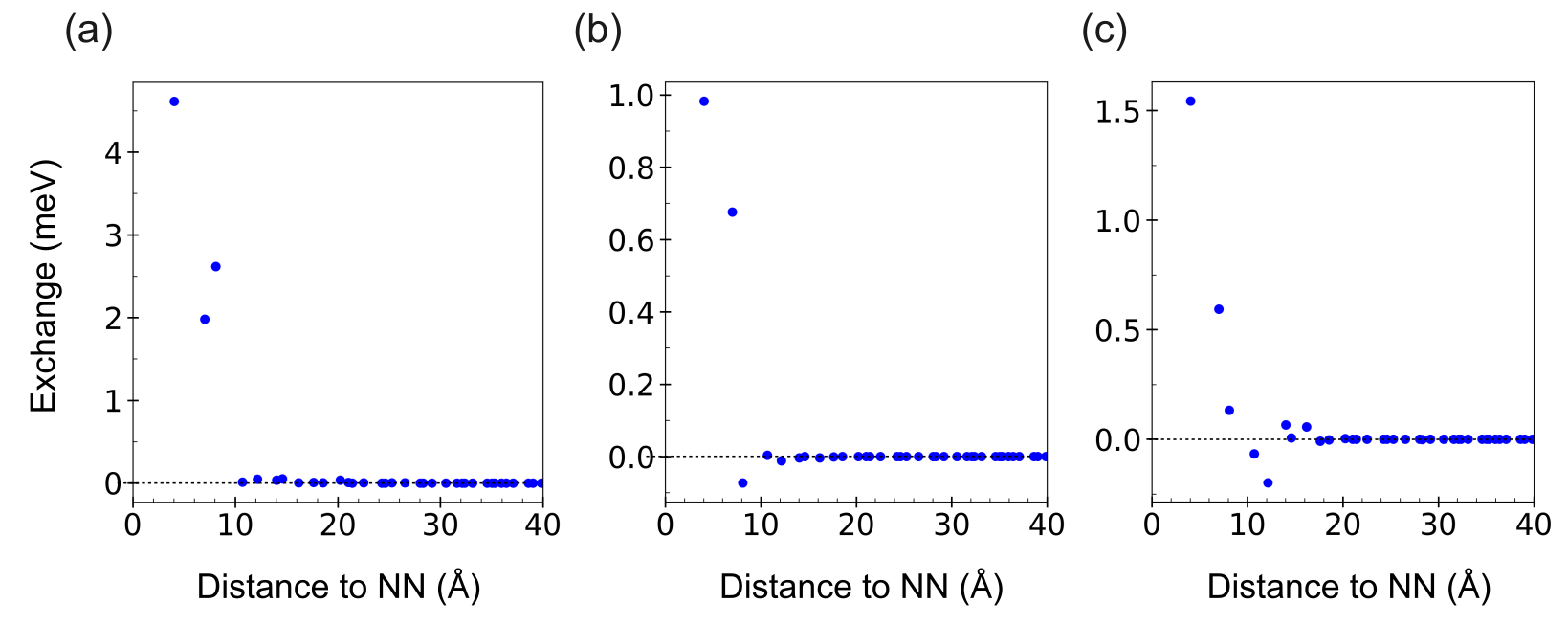}
    \caption{Exchange parameters dependent on nearest-neighbors distance for a) ${\rm CrI_3}$ doped with 1 holes/u.c., b) undoped system, and c) system doped with 0.5 electrons/u.c.}
    \label{fig:nn}
\end{figure}

In this section, we show that the Hamiltonian defined in equation (3) of the main text captures all essential NN interactions. Calculating exchange parameters beyond the third NN by energy mapping is computationally demanding, as it requires huge supercells. To evaluate isotropic exchange parameters for more distant neighbors, we employed the \texttt{TB2J} method. For both the undoped and doped system with 1 holes/u.c., the exchanges beyond the third NN are negligible, as illustrated in figure~\ref{fig:nn}(a) and (b). However, the fifth NN exchange in the doped system with 0.5 electrons/u.c. reaches the value of $0.22\,{\rm meV}$, as shown in the figure~\ref{fig:nn}(c). This is the largest value among all further-neighbor exchanges across the doping range considered. Since all exchanges beyond the third NN remain below $0.22,{\rm meV}$ for all doping concentrations, it is fully justified to restrict the model to first, second, and third NN interactions.

\section{Energy mapping method}
Here, we outline the energy mapping procedure used with the four-state method to obtain the parameters in Hamiltonian (3) of the main text. The element $\alpha\beta$ (e.g. $xz$) of the generalized exchange matrix $\mathcal{J}_{ij}^{\alpha\beta}$ is calculated between magnetic moments on sites $i$ and $j$. For this purpose, four configurations are generated by flipping the magnetic moments along these directions: $(\hat{\bf \alpha},\hat{\bf \beta}),\,(\hat{\bf \alpha},-\hat{\bf \beta}),\,(-\hat{\bf \alpha},\hat{\bf \beta}),\,(-\hat{\bf \alpha},-\hat{\bf \beta})$ which correspond to the energies $E_{\uparrow\uparrow},\, E_{\uparrow\downarrow},\, E_{\downarrow\uparrow},\, E_{\downarrow\downarrow}$, respectively.
The magnetic moments on all other sites in the supercell are constrained to be orthogonal to both $\hat{\bf \alpha}$ and $\hat{\bf \beta}$, and kept in the same direction for all four calculations. Assuming that all the magnetic moments in the system are with magnitude $S$ ($S=3/2$ in CrI$_3$), the element of an exchange matrix $\mathcal{J}_{ij}$ is obtained as
\begin{equation}
    \mathcal{J}_{ij}^{\alpha\beta} = \frac{E_{\uparrow\uparrow}+E_{\downarrow\downarrow}-E_{\uparrow\downarrow}-E_{\downarrow\uparrow}}{4S^2}.
\end{equation}
The single-ion anisotropy (SIA) parameter $A_{i,zz}$ on site $i$ is computed in a similar way. Concretely, the magnetic moment on site $i$ is constrained along four directions: $\hat{\bf z},\,-\hat{\bf z},\hat{\bf x}$ and $-\hat{\bf x}$, producing the energies $E_1,\,E_2,\,E_3,\,E_4$, respectively. Other magnetic moments in the supercell are constrained along $\hat{y}$. The energy mapping for SIA parameter is given by
\begin{equation}
    A_{i,zz} = \frac{E_1+E_2-E_3-E_4}{2S^2}.
\end{equation}

Exchange tensors were calculated between atomic site pairs depicted in figure 1(a) of the main text. Exchange matrices for all doping concentrations are listed in table~\ref{tab:J1}, while the SIA and MAE values are provided in table~\ref{tab:sia-mae}.

\begin{table}
\centering
\caption{Exchange matrix parameters of the first, second, and third NN for all doping concentrations.}
\begin{tabular}{|c|c|rrrrrrrrr|}
\hline
NN exchange&conc. & $J_{xx}$ & $J_{yy}$ & $J_{zz}$ & $J_{xy}$ & $J_{yx}$ & $J_{xz}$ & $J_{zx}$ & $J_{yz}$ & $J_{zy}$\\
\hline
\multirow{11}{*}{$\mathcal{J}_1$} 
&-1.000 &6.11 & 5.87 & 7.36 & -0.22 & -0.22 & -0.32 & -0.32 & -0.53 & -0.53\\
&-0.750 &4.74 & 4.43 & 5.81 & -0.29 & -0.29 & -0.27 & -0.27 & -0.44 & -0.44\\
&-0.500 &3.89 & 3.50 & 4.52 & -0.36 & -0.36 & -0.23 & -0.23 & -0.39 & -0.39\\
&-0.375 &3.56 & 3.15 & 4.02 & -0.38 & -0.38 & -0.22 & -0.22 & -0.37 & -0.37\\
&-0.250 &3.31 & 2.88 & 3.65 & -0.38 & -0.38 & -0.20 & -0.20 & -0.35 & -0.35\\
&-0.125 &3.11 & 2.70 & 3.35 & -0.41 & -0.41 & -0.21 & -0.21 & -0.33 & -0.33\\
&0.000 &2.97 & 2.54 & 2.82 & -0.36 & -0.36 & -0.21 & -0.21 & -0.38 & -0.38\\
&0.125 &2.92 & 2.83 & 2.69 & -0.08 & -0.08 & -0.07 & -0.07 & -0.10 & -0.10\\
&0.250 &2.88 & 3.08 & 2.68 & 0.19 & 0.19 & 0.05 & 0.05 & 0.19 & 0.19\\
&0.375 &2.98 & 3.39 & 2.72 & 0.40 & 0.40 & 0.17 & 0.17 & 0.39 & 0.39\\
&0.500 &3.10 & 3.76 & 2.80 & 0.58 & 0.58 & 0.29 & 0.29 & 0.56 & 0.56\\
\hline
\multirow{11}{*}{$\mathcal{J}_2$} 

&-1.000 &1.80 & 1.76 & 1.89 & 1.96 & -1.96 & -0.05 & -0.05 & -0.44 & 0.44\\
&-0.750 &1.15 & 1.13 & 1.22 & 1.49 & -1.49 & -0.01 & -0.01 & -0.30 & 0.30\\
&-0.500 &0.75 & 0.79 & 0.88 & 0.82 & -0.82 & -0.01 & -0.01 & -0.18 & 0.18\\
&-0.375 &0.62 & 0.68 & 0.79 & 0.54 & -0.54 & -0.00 & -0.00 & -0.13 & 0.13\\
&-0.250 &0.52 & 0.59 & 0.73 & 0.32 & -0.32 & 0.01 & 0.01 & -0.13 & 0.13\\
&-0.125 &0.49 & 0.57 & 0.62 & 0.14 & -0.14 & -0.01 & -0.01 & -0.08 & 0.08\\
&0.000 &0.61 & 0.70 & 0.59 & -0.09 & 0.09 & 0.03 & 0.03 & -0.09 & 0.09\\
&0.125 &0.69 & 0.49 & 0.51 & 0.02 & -0.02 & 0.02 & 0.02 & -0.01 & 0.01\\
&0.250 &0.74 & 0.41 & 0.38 & 0.08 & -0.08 & -0.02 & -0.02 & -0.01 & 0.01\\
&0.375 &0.79 & 0.34 & 0.32 & 0.13 & -0.13 & -0.03 & -0.03 & -0.02 & 0.02\\
&0.500 &0.81 & 0.32 & 0.28 & 0.19 & -0.19 & -0.02 & -0.02 & -0.03 & 0.03\\
\hline
\multirow{11}{*}{$\mathcal{J}_3$} 
&-1.000 &2.84 & 2.89 & 2.11 & -0.07 & -0.07 & 0.01 & 0.01 & -0.02 & -0.02\\
&-0.750 &2.12 & 2.14 & 1.50 & -0.04 & -0.04 & -0.00 & -0.00 & -0.00 & -0.00\\
&-0.500 &1.20 & 1.19 & 0.97 & -0.02 & -0.02 & -0.00 & -0.00 & 0.01 & 0.01\\
&-0.375 &0.79 & 0.79 & 0.72 & -0.01 & -0.01 & -0.00 & -0.00 & 0.01 & 0.01\\
&-0.250 &0.38 & 0.38 & 0.49 & -0.00 & -0.00 & 0.02 & 0.02 & -0.01 & -0.01\\
&-0.125 &0.05 & 0.05 & 0.20 & -0.00 & -0.00 & 0.07 & 0.07 & 0.01 & 0.01\\
&0.000 &-0.11 & -0.09 & -0.10 & 0.11 & 0.11 & -0.01 & -0.01 & 0.01 & 0.01\\
&0.125 &-0.05 & -0.04 & 0.03 & -0.01 & -0.01 & 0.05 & 0.05 & 0.04 & 0.04\\
&0.250 &0.01 & 0.04 & 0.13 & -0.02 & -0.02 & -0.00 & -0.00 & 0.02 & 0.02\\
&0.375 &0.06 & 0.06 & 0.22 & 0.01 & 0.01 & -0.02 & -0.02 & 0.03 & 0.03\\
&0.500 &0.12 & 0.11 & 0.29 & 0.01 & 0.01 & -0.03 & -0.03 & 0.03 & 0.03\\
\hline
\end{tabular}
\label{tab:J1}
\end{table}

\begin{table}
\centering
\caption{SIA and MAE parameters for all doping concentrations.}
\begin{tabular}{|c|rrrrrrrrrrr|}
\hline 
Conc. & -1.000 & -0.750 & -0.500 & -0.375 & -0.250 & -0.125 & 0.000 & 0.125 & 0.250 & 0.375 & 0.500\\
\hline
SIA & -0.38 & -0.34 & -0.16 & -0.03 & 0.10 & -0.03 & 0.29 & 0.24 & 0.39 & 0.36 & 0.36\\
MAE & 3.49 & 4.32 & 4.43 & 4.32 & 3.96 & 3.19 & 0.53 & -0.16 & -0.97 & -1.75 & -2.29\\
\hline
\end{tabular}
\label{tab:sia-mae}
\end{table}

\end{document}